\documentclass[a4paper,11pt]{article}
\pdfoutput=1 % if your are submitting a pdflatex (i.e. if you have
\usepackage{jinstpub} % for details on the use of the package, please
\usepackage{subfig}
\usepackage{wrapfig}

\title{\boldmath Optimization of RPCs read-out panel with electromagnetic simulation.}

\author[a,b,1]{E. Alunno Camelia,}
\author[b,1]{R. Cardarelli,}

\author[a,b]{A. Di Ciaccio,}
\author[a,b]{S. Bruno,}
\author[a,b]{A. Caltabiano,}
\author[b]{R. Camarri,}
\author[b]{B. Liberti,}
\author[a,b]{L. Massa,}
\author[a,b]{L. Pizzimento,}
\author[a,b]{A. Rocchi}

\affiliation[a]{University of Rome TorVergata  }%Via della ricerca scientifica 1 , Roma, Italia}
\affiliation[b]{INFN Roma 2\\Via della ricerca scientifica 1 , Roma, Italia}

\emailAdd{elio.alunnocamelia@roma2.infn.it}

\abstract{ With the upgrade of the RPCs [1]-[2] and the increase of its performances, the study and the optimization of the read-out panel is necessary in order to maintain the signal integrity and to reduce the intrinsic crosstalk. Through Electromagnetic Simulation, performed with CST Studio Suite, new panels design are tested and their crosstalk property are studied. The behavior of different type of panel is shown, in particular a panel with the decoupling strip connected through their characteristic impedance to the ground plane is simulated.
\begin{figure}[htbp!]
\centering
\includegraphics[scale=0.5]{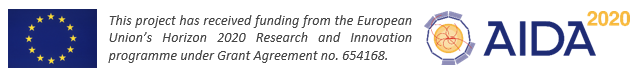}
\hspace{6mm}
\includegraphics[scale=0.2]{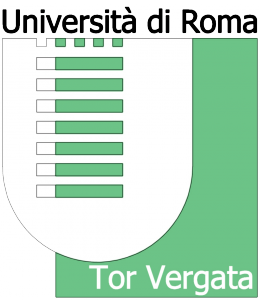}
\hspace{6mm}
\includegraphics[scale=0.5]{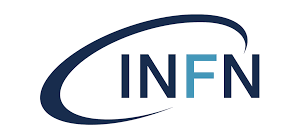}
\end{figure}
}

\proceeding{XIV RPC 2018 - Workshop on resistive plate chambers and related detectors\\
  19-23 February 2018\\
  Puerto Vallarta, Jalisco State, MEXICO}

\begin{document}
\maketitle
\flushbottom
\section{Crosstalk Simulation with CST STUDIO}
\begin{figure}[!h]
\centering
\subfloat[][\small{Floating model}]
{\includegraphics[trim=0.3cm 0.1cm 0.3cm 0.3cm, clip=true, width=0.35\textwidth]{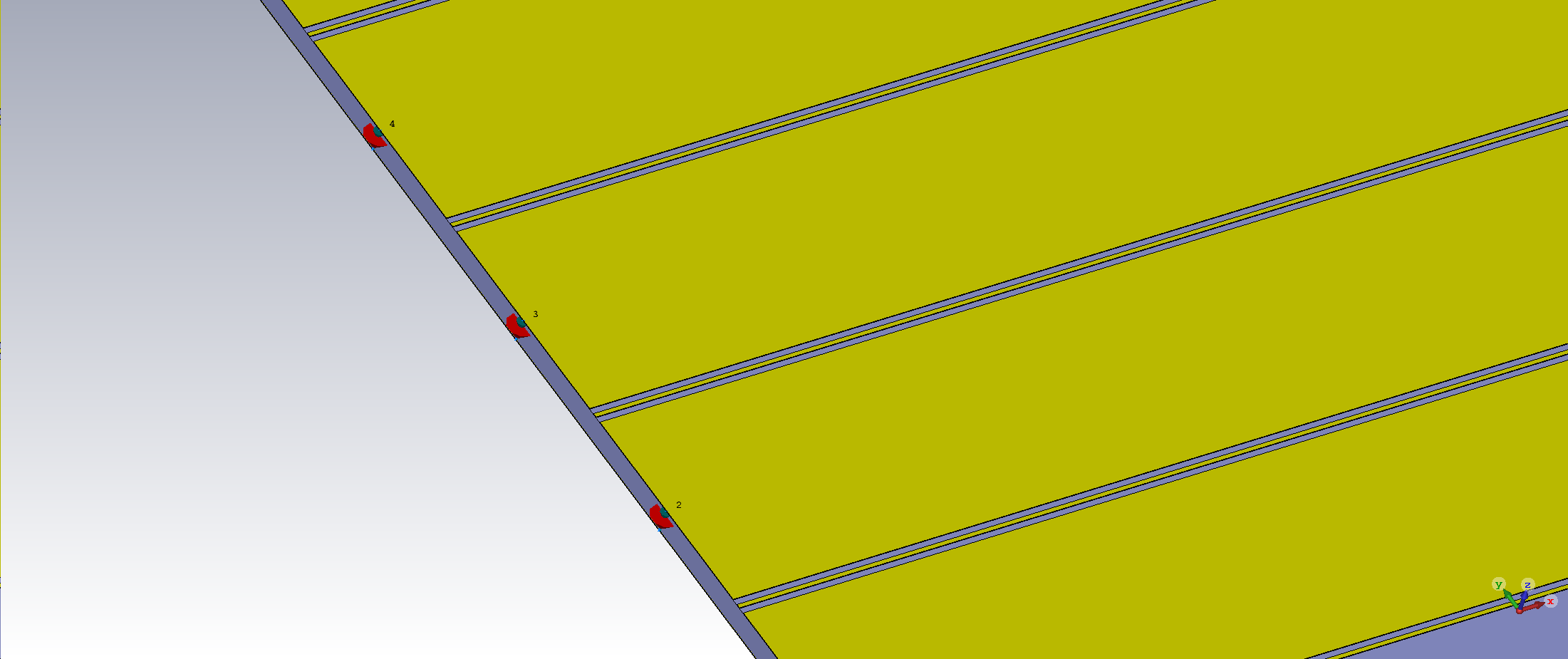}} \quad
\subfloat[][\small{Grounded model}]
{\includegraphics[trim=0.3cm 0.1cm 0.3cm 0.3cm, clip=true,width=0.32\textwidth]{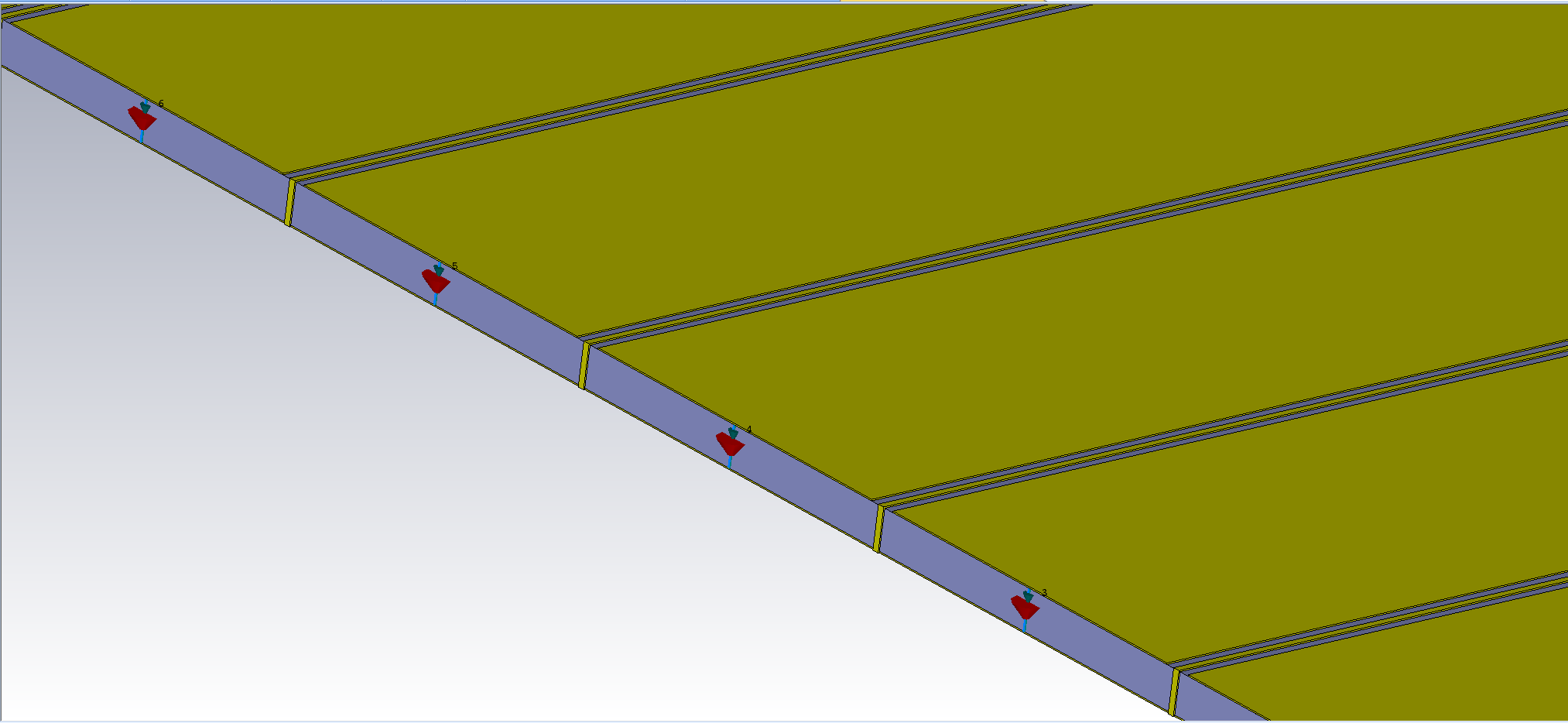}} \\
\subfloat[][\small{NoGround model}]
{\includegraphics[trim=0.3cm 0.25cm 0.3cm 0.3cm, clip=true,width=0.35\textwidth]{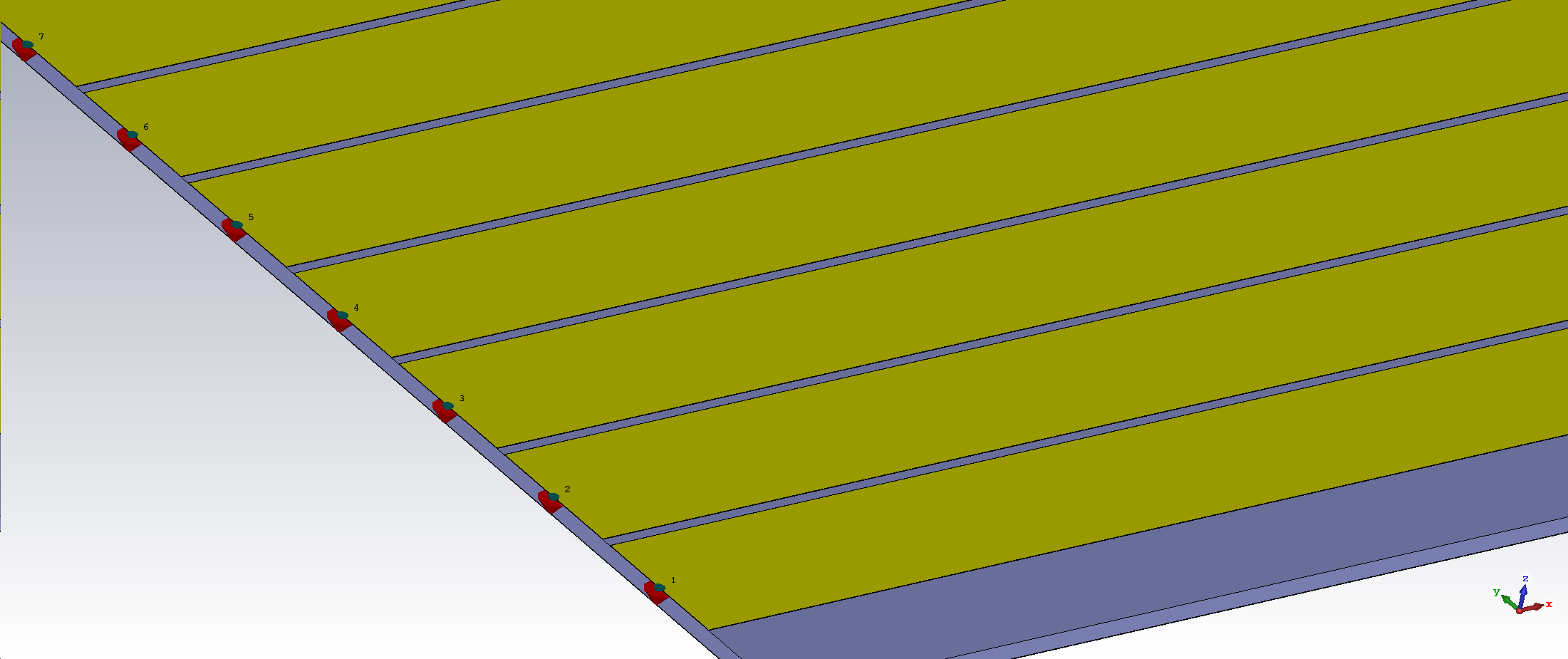}} \quad
\subfloat[][\small{Adapted model}]
{\includegraphics[trim=0.3cm 0.1cm 0.3cm 0.3cm, clip=true, width=0.32\textwidth]{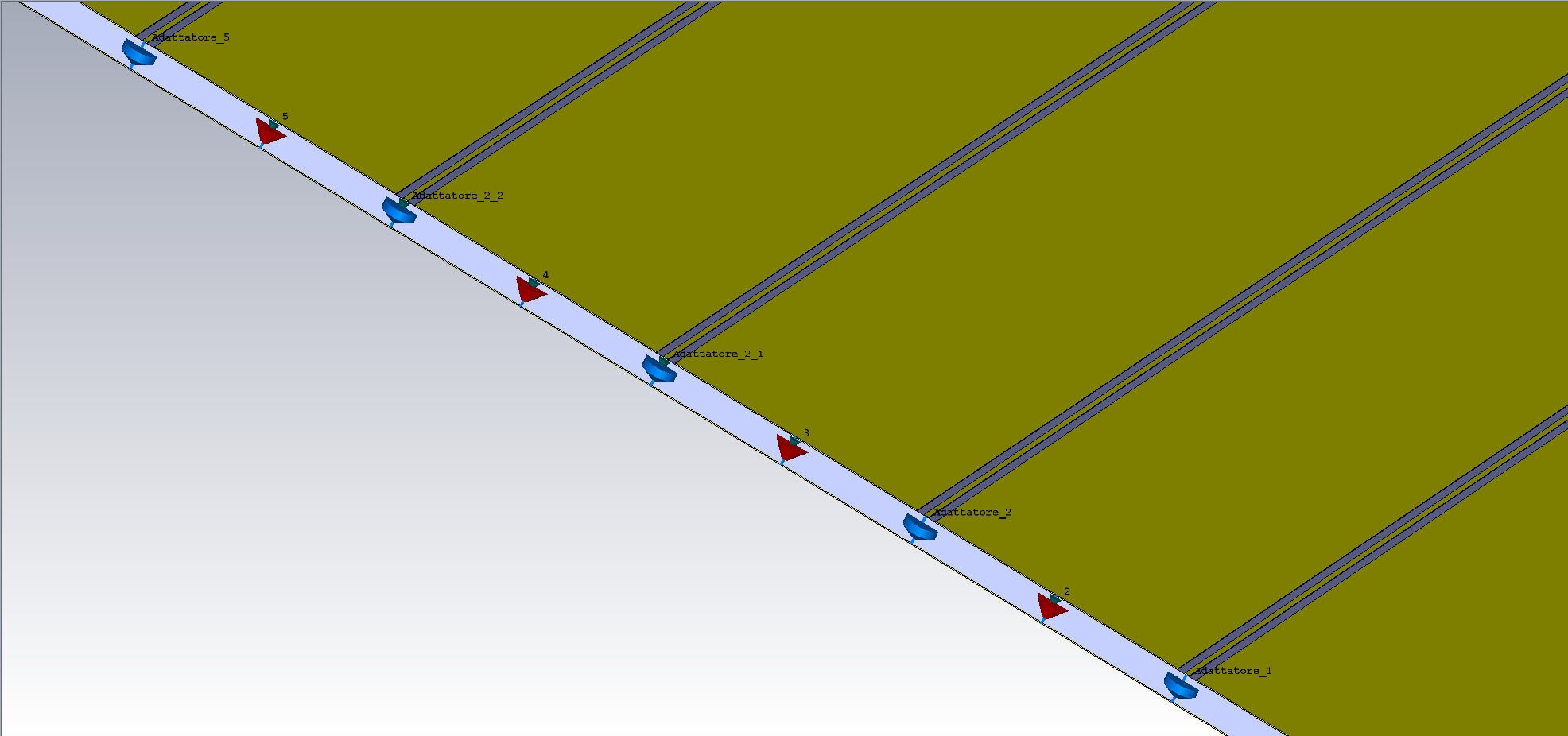}}
\caption{The four different simulated models}
\label{fig:models}
\end{figure}

The subjects of these simulations are four different configuration of the pick-up electrode of the RPC detector. 
In figure \ref{fig:models} it can see the simulated models, that differ from each other by the presence or not of a decoupling microstrip between the signal strip and by their termination. 
The "noGround" model is a simple readout panel with only ground plane and strip; the second is "Grounded" in which there is a decoupling microstrip connected directly to the ground plane; in the "floating" model the decoupling microstrip is unconnected and finally in the "adapted" model these last are adapted to their characteristic impedance.
The simulation was performed with CST STUDIO [3], an electromagnetic simulation program, that solves the integral form of the Maxwell equations.
Every strip is connected to a signal generator to emulate the signal in the RPC gap, on the other side every strip is adapted to its characteristic impedance Z $\sim 23 \Omega.$ 
At first, the simulation was performed by exciting a single generator at a time, then with all the generators together, to simulate one real situation of multi-crosstalk between all the strips. As a consequence, the pure signal of crosstalk turned out to be the subtraction of the two signals of the two different simulations. 
These simulations have been performed for different values of the signal distribution width within the gap, to simulate various pitch dimensions (without changing the pitch on the CAD model).

\pagebreak

\subsection{Simulation Results}

\begin{figure}[!h]
\centering
\subfloat[][\small{Crosstalk signal for 7 mm sigma distribution}]
{\includegraphics[trim=0.3cm 0.1cm 0.3cm 0.3cm, clip=true, width=0.35\textwidth]{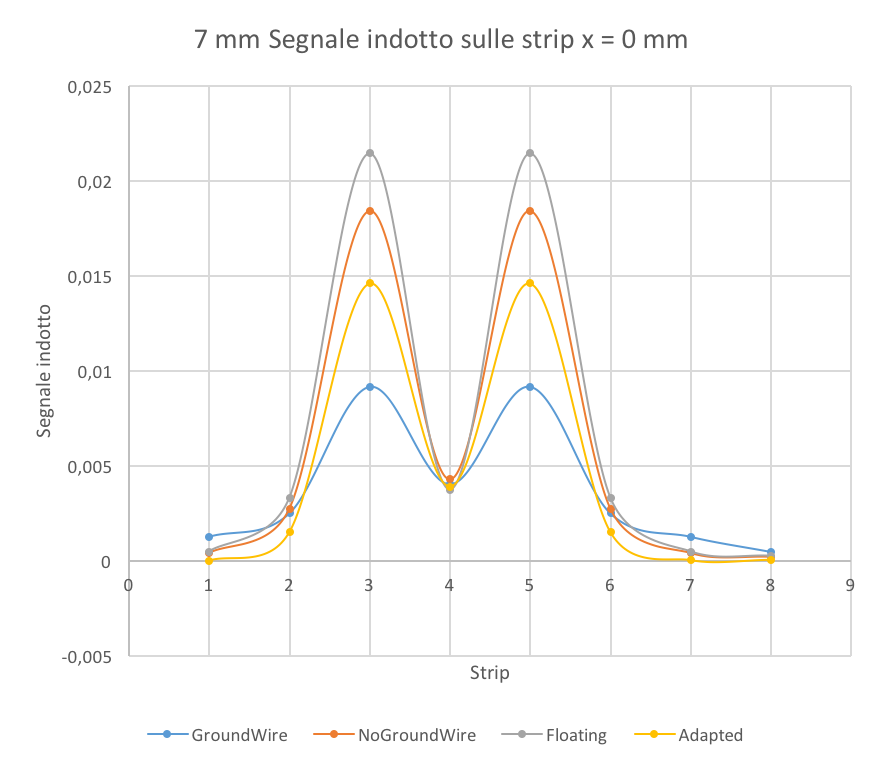}} \quad
\subfloat[][\small{Crosstalk signal for 13 mm sigma distribution}]
{\includegraphics[trim=0.3cm 0.1cm 0.3cm 0.3cm, clip=true,width=0.35\textwidth]{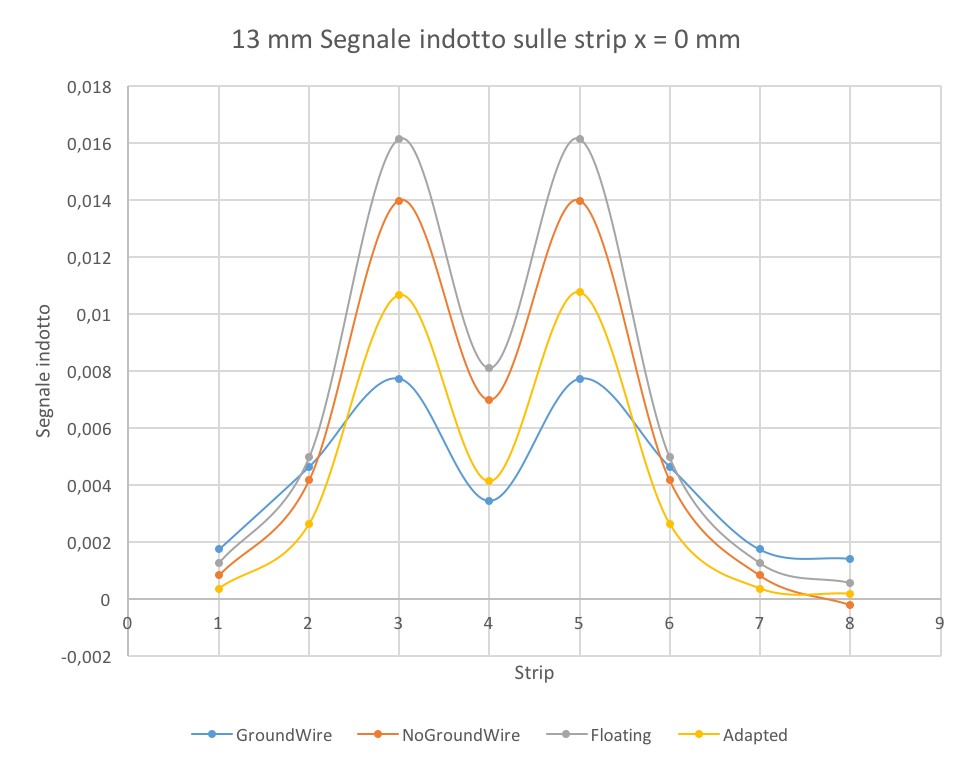}} \\
\subfloat[][\small{Crosstalk signal for 26 mm sigma distribution}]
{\includegraphics[trim=0.3cm 0.25cm 0.3cm 0.3cm, clip=true,width=0.35\textwidth]{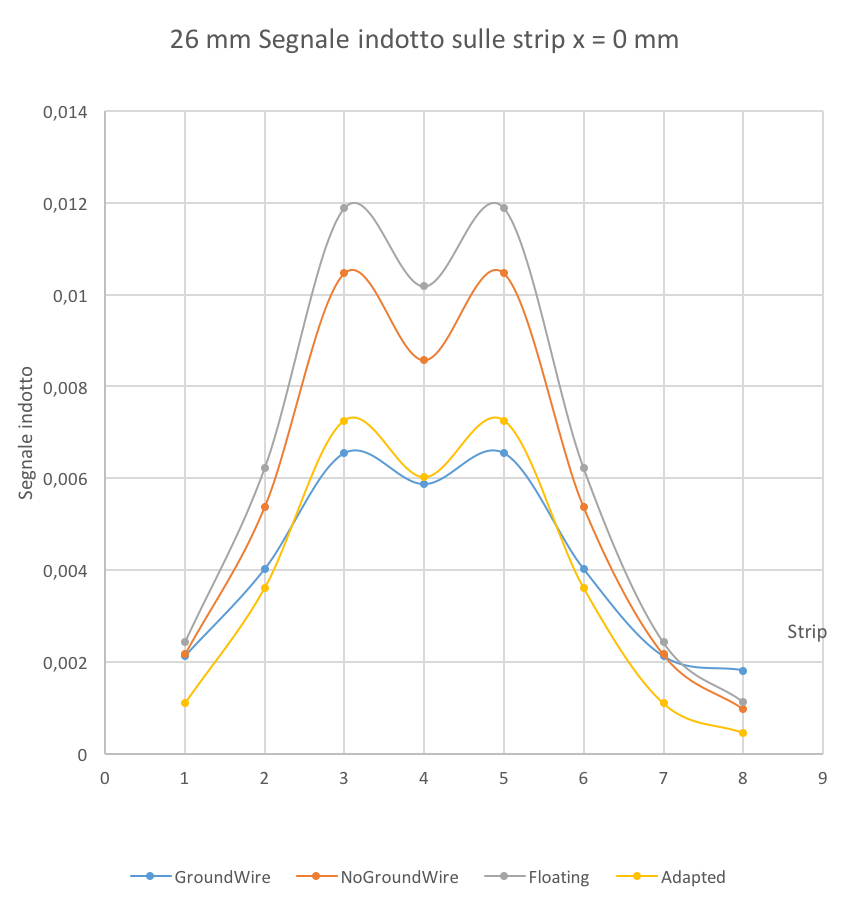}} \quad
\subfloat[][\small{Crosstalk signal for 52 mm sigma distribution}]
{\includegraphics[trim=0.3cm 0.1cm 0.3cm 0.3cm, clip=true, width=0.35\textwidth]{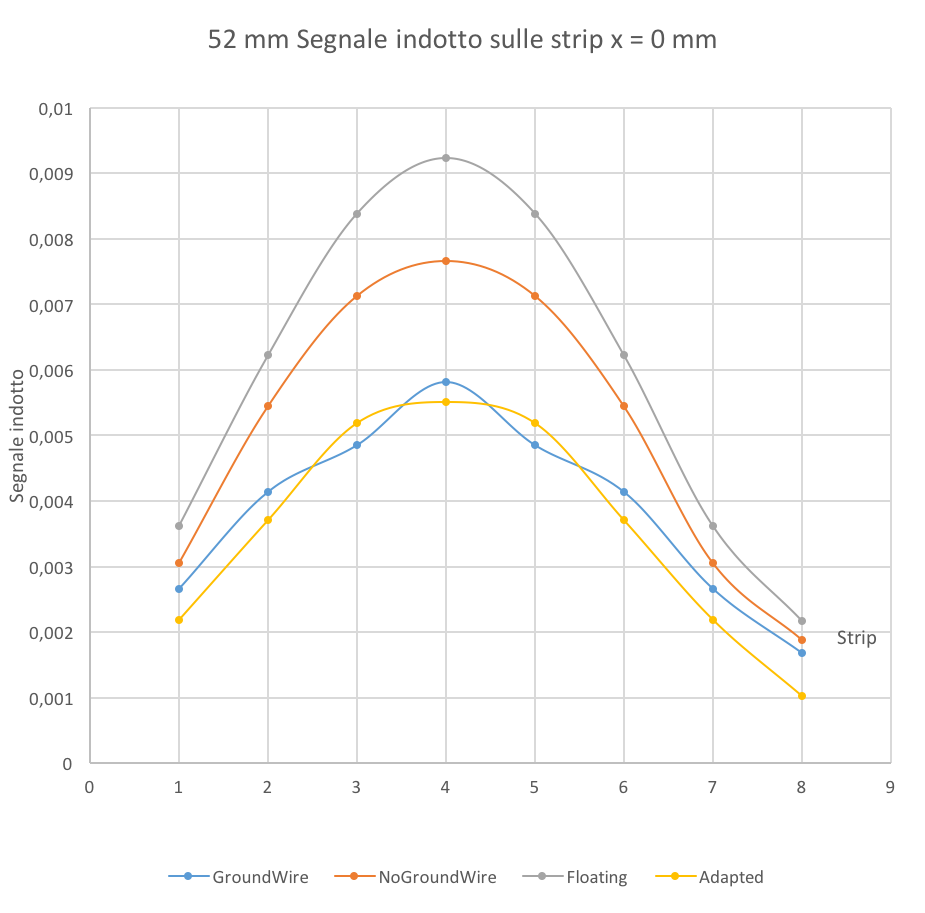}}
\caption{The four different simulated models}
\label{fig:graphics}
\end{figure}

\begin{figure}[!h]
\centering
\subfloat[][\small{NogroundWire model}]
{\includegraphics[trim=0.3cm 0.1cm 0.3cm 0.3cm, clip=true, width=0.46\textwidth]{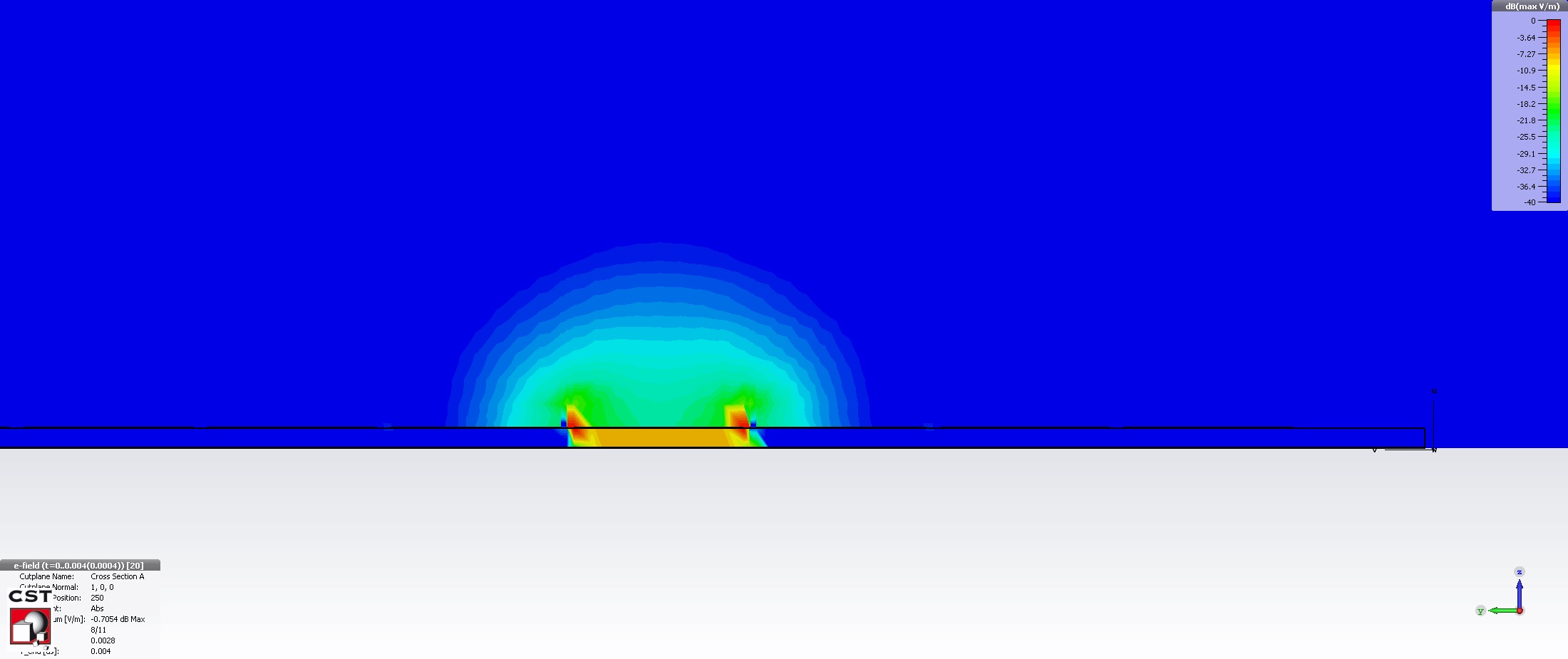}} \quad
\subfloat[][\small{adapted model}]
{\includegraphics[trim=0.3cm 0.1cm 0.3cm 0.3cm, clip=true, width=0.46\textwidth]{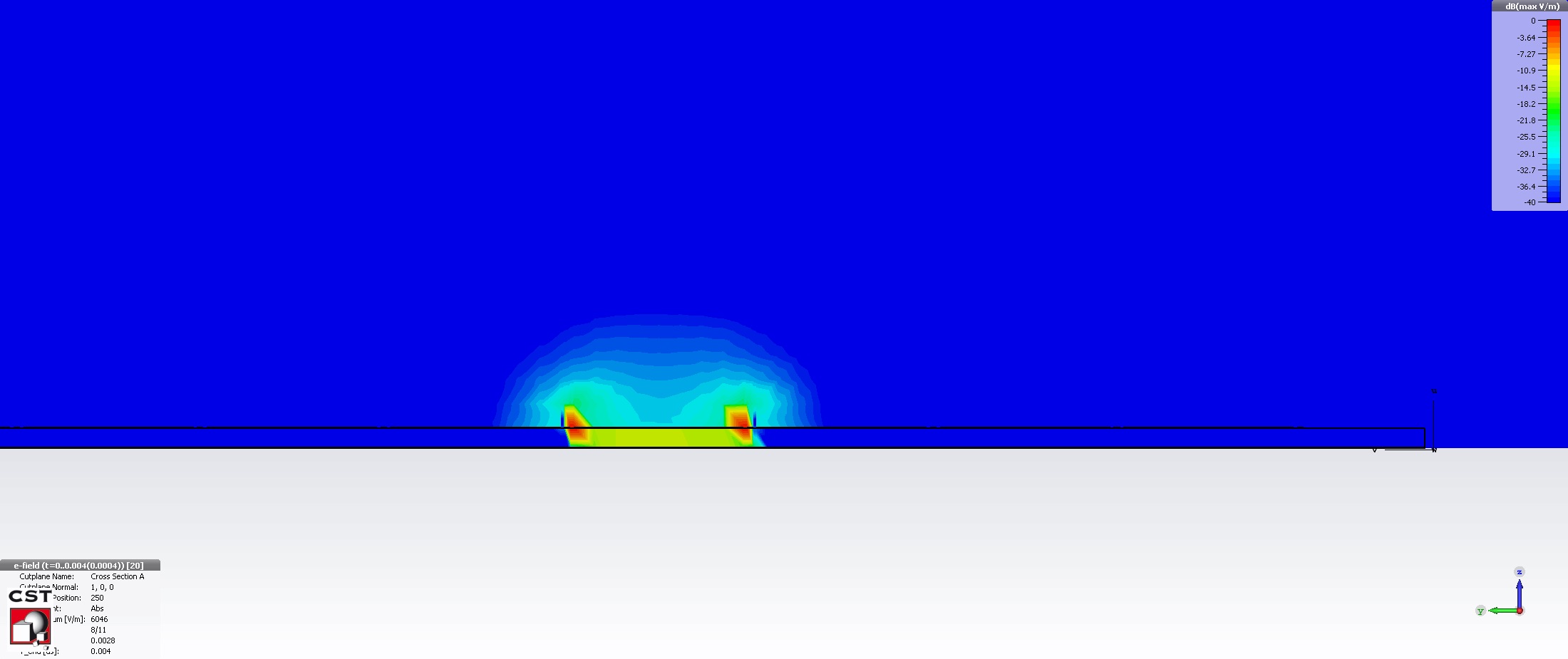}}
\caption{Electric field on a strip transversal plane}
\label{fig:Efield}
\end{figure}

From the result in the figure \ref{fig:graphics}, is clear that the floating and the grounded model are the worst in term of induced signal, in all distributions, as regards the signal on the neighboring prime strips, the grounded model has an intrinsic crosstalk significantly lower than the NoGroundWire and Floating, while compared to the Adapted model, the gap decreases. It is noted, in effects, that the Adapted model is advantageous in terms of crosstalk regarding the neighboring second strips. This improvement could be very interesting if the detector threshold is lowered. If so, with a bigger distribution (or a decrease in the pitch) one would work in a cluster-size situation equal to three, in almost all cases. The adoption of the Adapted panel would solve the problem of a possible increase of
 the cluster-size to five, thanks to better performance of this model in the neighboring second strips.
In figure \ref{fig:Efield} it is possible to see the electric field in the transverse plane with respect to the plane of the 
strips: on the left there is the NoGroundWire model, without the decoupling strip and the electric field
 lines arrive on the neighboring strip, while on the right there is the Adapted model, the electric field lines
are more confinated and this involves a lower crosstalk signal.

\newpage

\subsection{The graphite layer}
\begin{figure}[!h]
\centering
\includegraphics[trim=0.3cm 0.1cm 0.3cm 0.3cm, clip=true, width=0.46\textwidth]{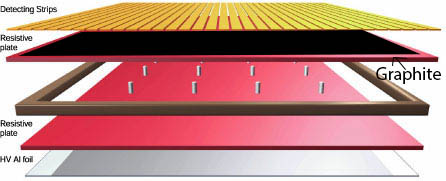}
\caption{RPC structure}
\label{fig:layers}
\end{figure}

In the figure \ref{fig:layers} we can see a tipical stratigraphy of a RPC detector, in which there is a thin layer of graphite to distribute the high voltage. This thin layer in theory is transparent to the induction signal and it should not affect the cluster size of the event. Instead what we observed is that the influence of the graphite layer is in relation with its resistivity.

\begin{figure}[!h]
\centering
\subfloat[][\small{Cluster size vs resistivity}]
{\includegraphics[trim=0.3cm 0.1cm 0.3cm 0.3cm, clip=true, width=0.5\textwidth]{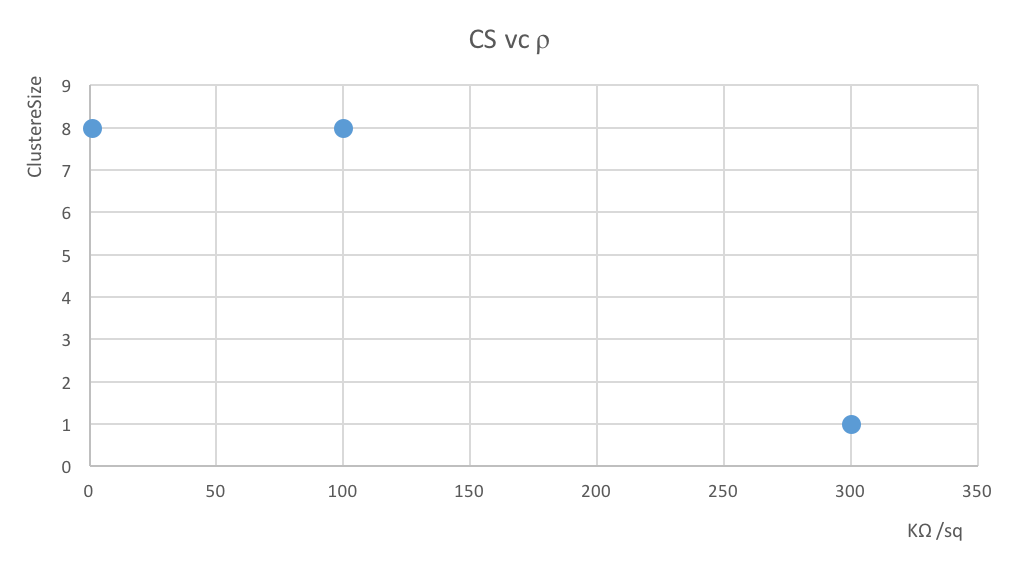}} \quad
\subfloat[][\small{Signal distributio with different material }]
{\includegraphics[trim=0.3cm 0.1cm 0.3cm 0.3cm, clip=true, width=0.4\textwidth]{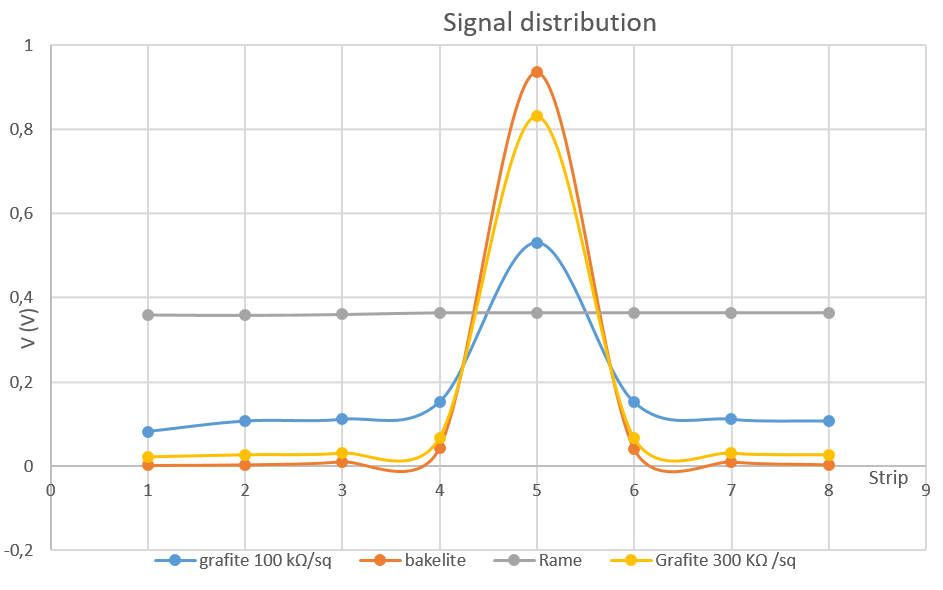}}
\caption{New layout design}
\label{fig:grap}
\end{figure}

For these reason the behavior of the panel was simulated in the presence of a thin layer of different material above it: graphite, bakelite and copper, varying the resistivity of the material and measuring the cluster-size of the event. As you can see in fig \ref{fig:grap}a the cluster-size decreases with increasing material resistivity, just as expected. With the graphite $\rho = 100 {𝑘\Omega}/{\Box}$ the situation is the same, in term of cluster size) as obtained with a conductive copper panel placed above the strip panel. Increasing the value of resistivity we return to a normal situation with CS = 1. In the figure \ref{fig:grap}b it is possible to see the signal distribution with the presence of different material above the strip panel. For resistivity value above $300 K\Omega / \Box$ we have the correct behavior.
\begin{itemize}
\item{Bakelite: Crosstalk signal on the first neighbor $\sim 6\%$ }
\item{Graphite $\rho = 100 {𝑘\Omega}/{\Box}$: Crosstalk signal on the first neighbor  $\sim 17 \%$}
\item{Graphite $\rho = 300 {𝑘\Omega}/{\Box}$: Crosstalk signal on the first neighbor  $\sim 8 \%$}
\end{itemize}

\section{New readout layout}
This is the new read-out electrode layout, designed starting from the simulation to minimize the crosstalk signal. 
The material is the fiberglass, used generally in the PCB construction. Every signal strip have a back-end and a front-end resistor of different value, and also every decoupling strip has this resistor to adapt his characteristic impedance, Z $\sim 100 \Omega$.
An innovation is the Front-end board integration, in fact the FE board and the panel strip are designed to perfectly match. Now the FE boars will be sold on the strip panel and not near the panel.
\begin{figure}[!h]
\centering
\subfloat[][\small{Strip design}]
{\includegraphics[trim=0.3cm 0.1cm 0.3cm 0.3cm, clip=true, width=0.46\textwidth]{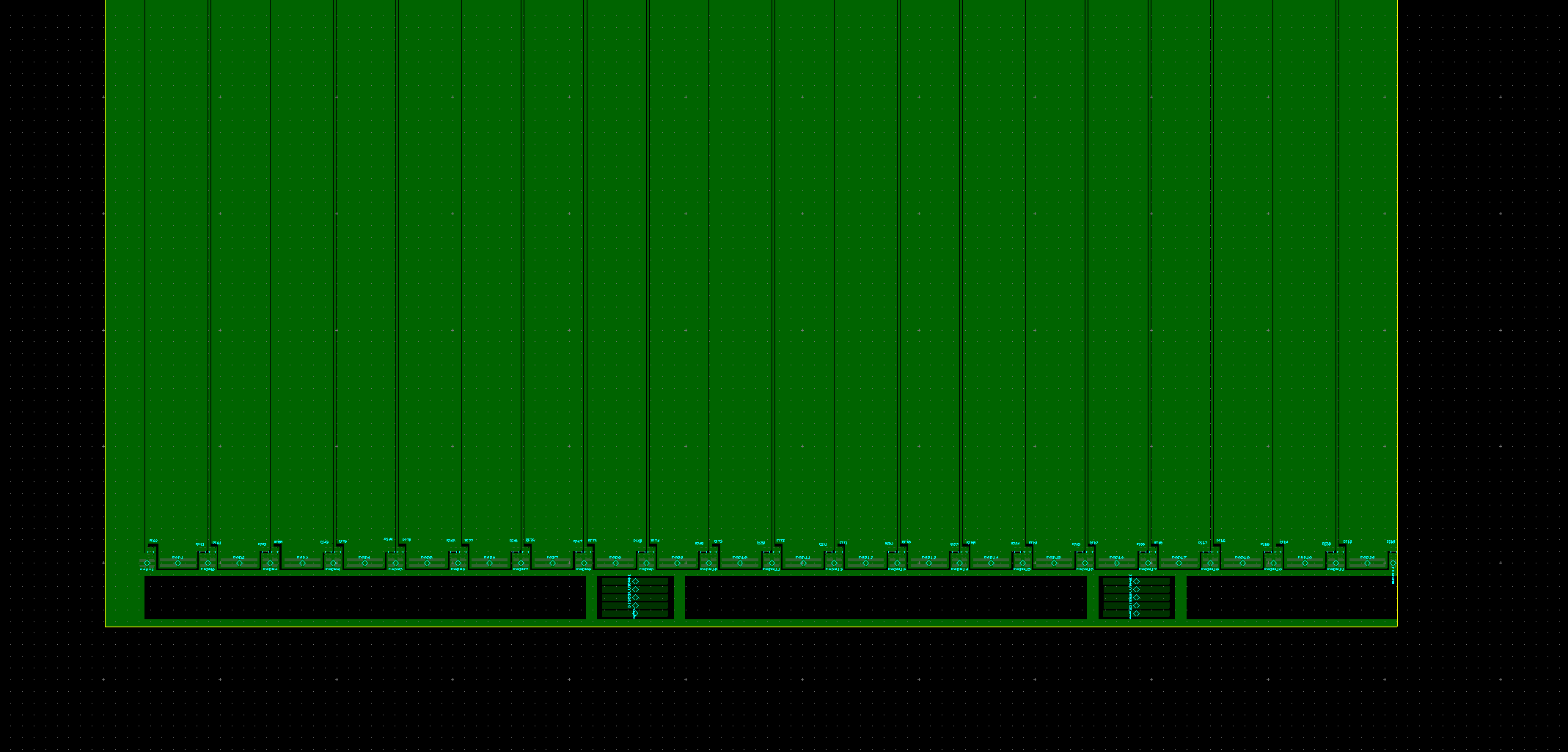}} \quad
\subfloat[][\small{Zoom }]
{\includegraphics[trim=0.3cm 0.1cm 0.3cm 0.3cm, clip=true, width=0.46\textwidth]{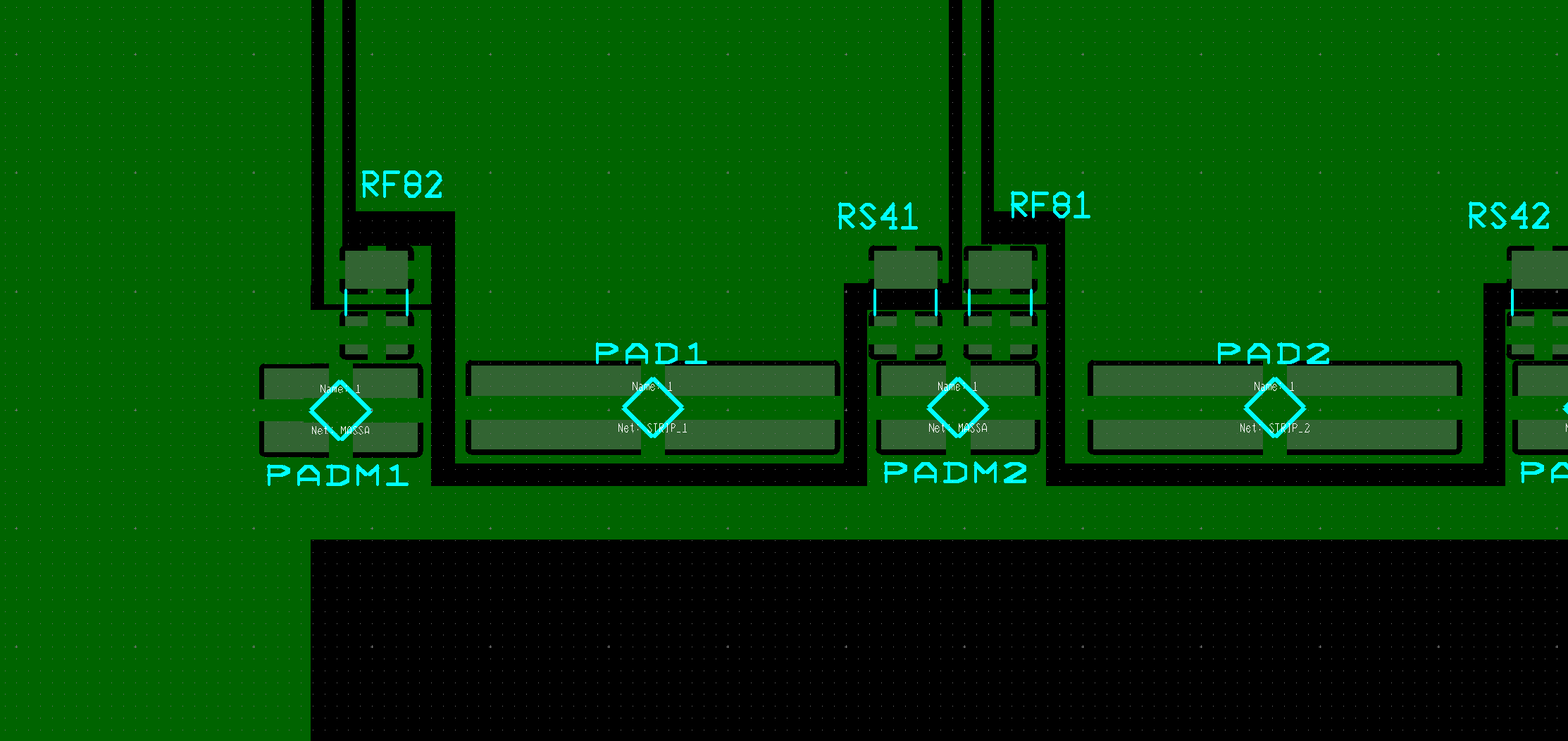}}
\caption{New layout design}
\label{fig:newlayout}
\end{figure}

\section{Test results}
\begin{figure}[!h]
\centering
\subfloat[][\small{New strip panel}]
{\includegraphics[trim=0.3cm 0.1cm 0.3cm 0.3cm, clip=true, width=0.34\textwidth]{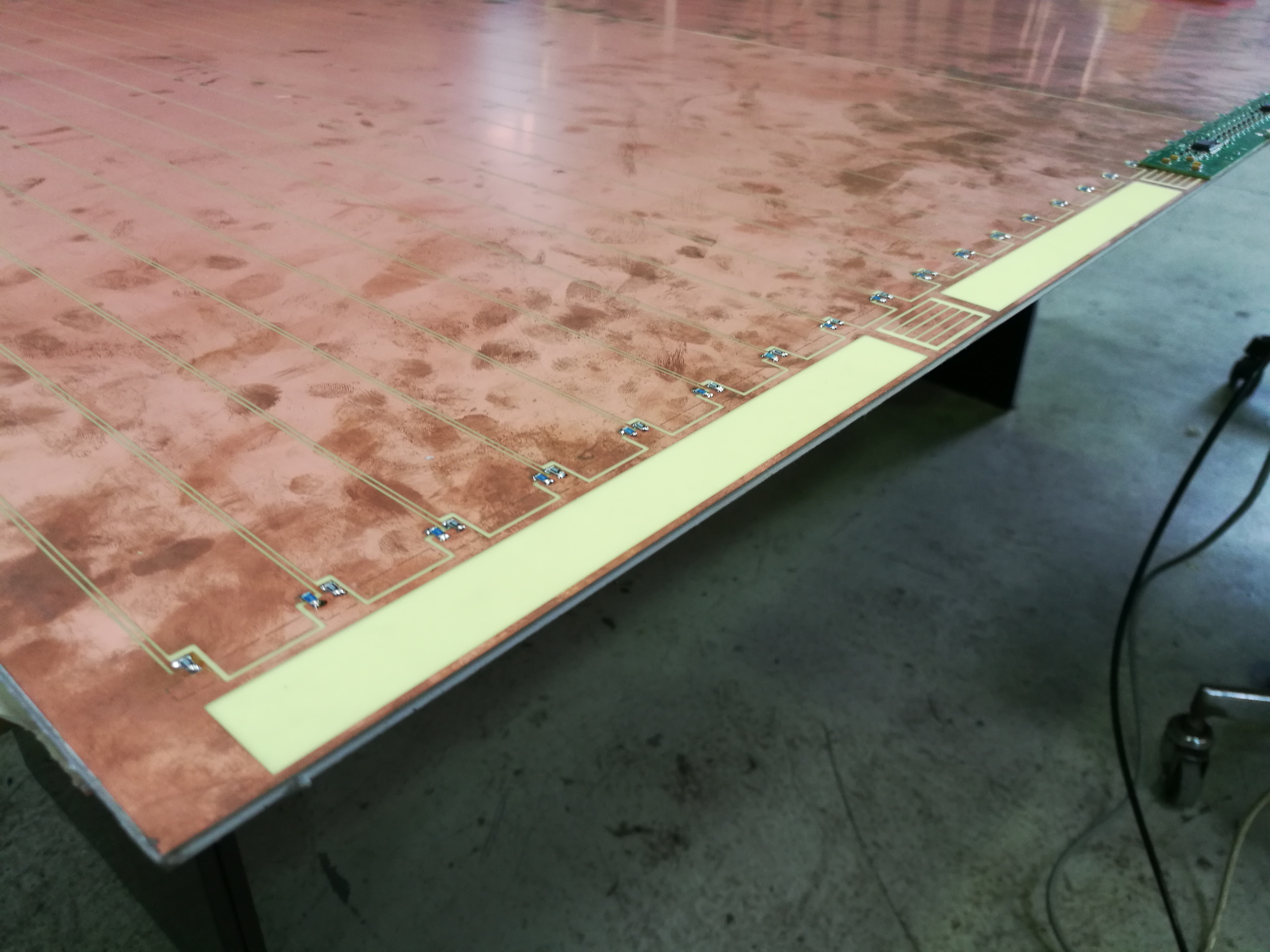}} \quad
\subfloat[][\small{New front end electronics}]
{\includegraphics[trim=0.3cm 0.1cm 0.3cm 0.3cm, clip=true, width=0.34\textwidth]{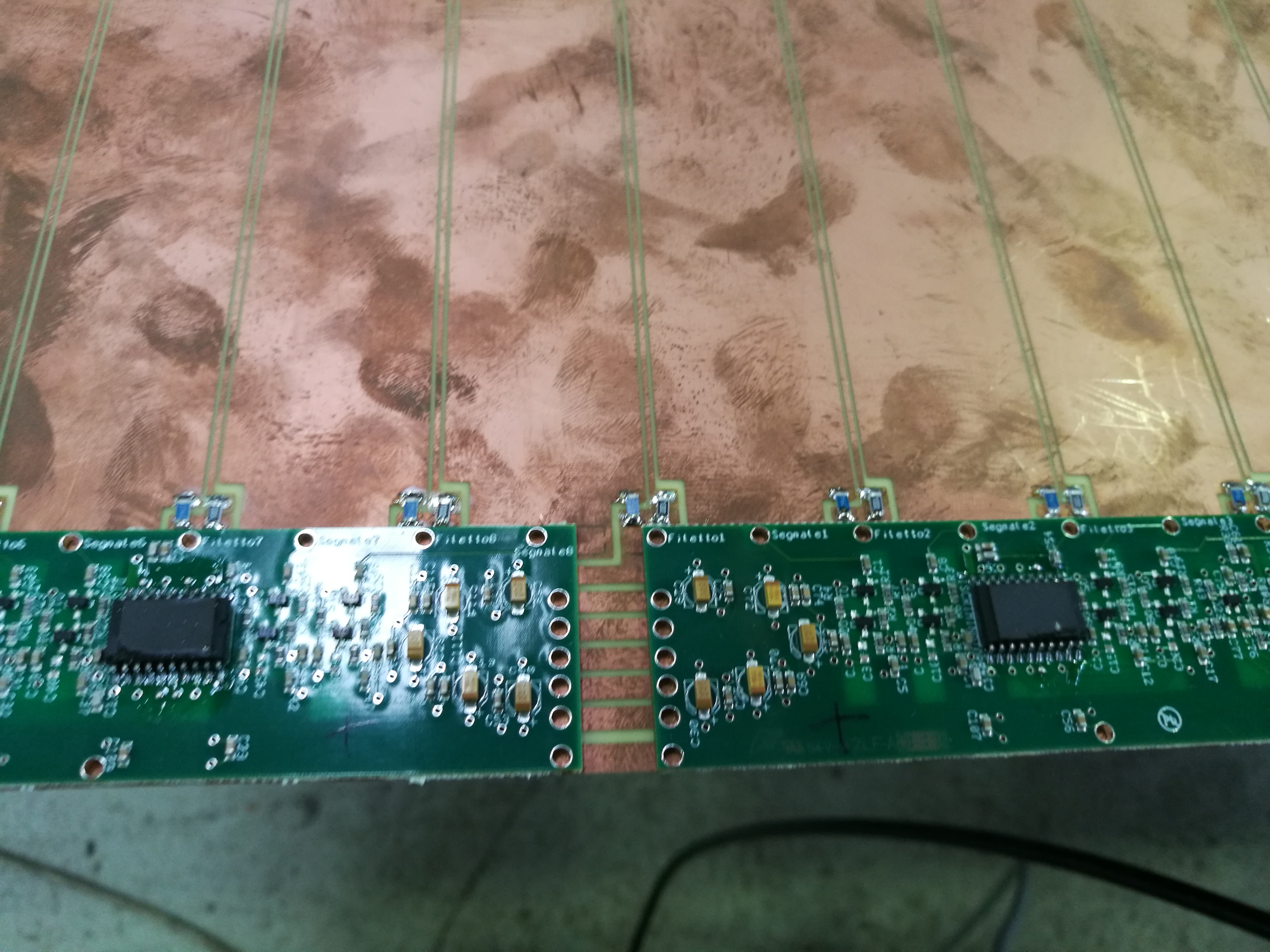}} \quad
\subfloat[][\small{Crosstalk signal in different condition}]
{\includegraphics[trim=0.3cm 0.1cm 0.3cm 0.3cm, clip=true, width=0.46\textwidth]{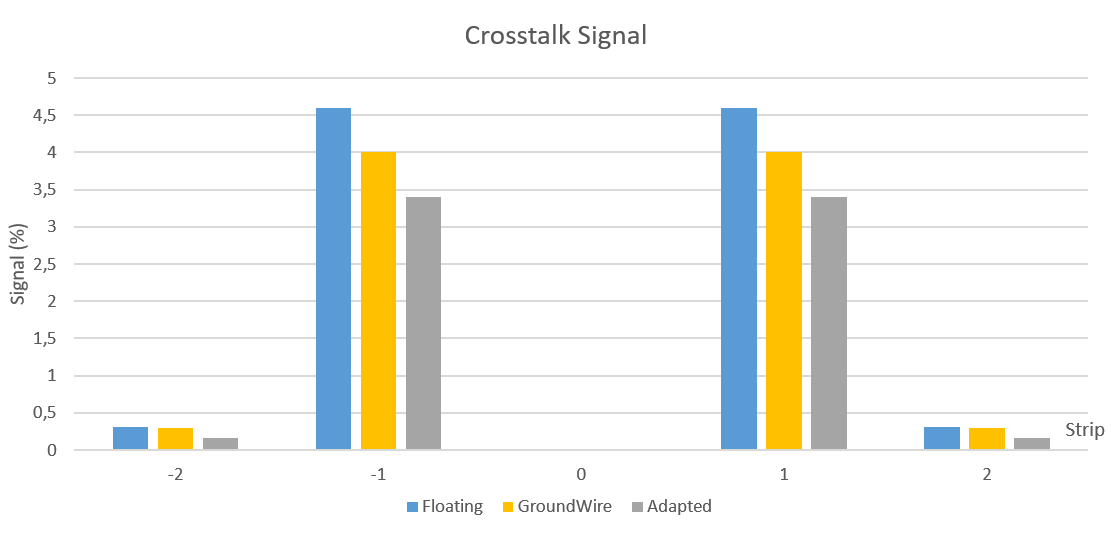}}
\caption{New layout design and crosstalk test results}
\label{fig:test}
\end{figure}

The figure \ref{fig:test} show the new strip panel and the new front-end electronics. From the reflectometry test the signal strip and decoupling strip impedance is in according to the simulation value.
It's possible to see in the graphic the measured crosstalk signal in three different situations: with the decoupling strip floating, adapted and directly connected to the ground plane.
The best situation, such as in the simulation, is the adapted one, in which the crosstalk signal is smaller in both the first and the second strip.

%\acknowledgments
%
%This is the most common positions for acknowledgments. A macro is
%available to maintain the same layout and spelling of the heading.
%\paragraph{Note added.} 

\pagebreak

\end{document}